\documentclass[12pt,onecolumn,tworows]{IEEEtran}

\usepackage{chemarrow}
\usepackage{CJK}
\usepackage{fancybox}
\usepackage{times}
\usepackage{amsmath}
\usepackage{amssymb}
\usepackage[dvips]{graphicx}
\usepackage{subfigure}
\usepackage{bm}
\usepackage{cite}
\usepackage{array}
\usepackage{algorithm}
\usepackage{algpseudocode}

\newcommand{\utwi}[1]{\mbox{\boldmath $ #1$}}

\begin{document}

 \baselineskip 25pt


\title{\Large\bf  Probability-constrained Power Optimization for Multiuser MISO Systems with Imperfect CSI: A Bernstein Approximation Approach}

\author{Weiqiang Xu,
~\IEEEmembership{Member,~IEEE},
Xiaodong~Wang,
~\IEEEmembership{Fellow,~IEEE},
Saleh Alshomrani

\thanks{W. Xu is with the School of Information Science \& Technology, Zhejiang Sci-Tech University, Hangzhou, 310018, P. R. China.
X. Wang is with the Electrical Engineering Department, Columbia University, New York,
NY, 10027, USA.
Email: wangx@ee.columbia.edu.
S. Alshomrani is with the Faculty of Computing \& Information Technology,
King Abdulaziz University, Jeddah, Saudi Arabia.Email: sshomrani@kau.edu.sa}
}

\maketitle

\begin{abstract}
We consider power allocations in downlink cellular wireless systems where the basestations are equipped with multiple
transmit antennas and the mobile users are equipped with single receive antennas. Such systems can be modeled
as multiuser MISO systems.
We assume that the multi-antenna transmitters employ some fixed beamformers to transmit data, and
the objective is to optimize the power allocation for different users to satisfy certain QoS
constraints, with imperfect transmitter-side channel state information (CSI).
Specifically, for MISO interference channels, we consider the
transmit power minimization problem and the max-min SINR problem. For MISO broadcast
channels, we consider the MSE-constrained transmit power minimization problem.
All these problems are formulated as probability-constrained optimization
problems. We make use of  the Bernstein approximation to conservatively
transform the probabilistic constraints into deterministic ones, and
consequently convert the original stochastic optimization problems into
convex optimization problems. However, the transformed problems cannot be straightforwardly solved using
standard solver, since one of the constraints is itself an
optimization problem.
We employ the
long-step logarithmic barrier cutting plane (LLBCP) algorithm
to overcome difficulty. Extensive simulation
results are provided to demonstrate the effectiveness of the proposed method, and the performance advantage over some existing methods.
\end{abstract}

\newpage
\section{Introduction}
\label{sec:introduction}

The multiuser multiple-input single-output (MISO) system can be used to model a communication
system where there is an asymmetry between the transmitters and the receivers in terms of the
number of antennas employed. For example, in a cellular system, typically the basestation  can
be equipped with multiple antennas, whereas the mobile users are equipped with single-antennas.
Then for the downlink transmission, when a single cell is considered, we have a MISO broadcast
channel; whereas when multiple cells are considered, we have a MISO interference channel.
Such multiuser MISO channels have been extensively studied in the literature.
In particular,   the achievable rate regions of the MISO interference channels have been investigated
in \cite{Shafiee} \cite{Jorswieck2008}. In \cite{Ivrlac} \cite{Castaneda}, the problem of maximizing the  sum rate of the
MISO broadcast channel is treated. The same problem is also considered in \cite{Song}, with the additional  max-min fairness constraint.
Moreover,  bargaining based game-theoretic solutions are given in \cite{Larsson}  for a two-user MISO interference channel,
  and in \cite{Nokleby} for the general  $K$-user  case. All these works assume that perfect channel state information (CSI) is available at the transmitters.

However, in practice, due to various reasons, such as
  estimation/quantization errors, delayed estimation, and limited feedback rate,
the assumption of  perfect CSI at the transmitter is unrealistic.
In the case of imperfect CSI,
the naive approach of treating the   imperfect CSI as
if it was perfect gives rise to non-robust design, which leads to
rapid   performance degradation   with the increasing error levels \cite{Jindal}.
Motivated by this fact,
a number of efforts have addressed robust approaches that can cope with
uncertainties in the channel knowledge.
Convex optimization techniques are important tools for
obtaining computationally efficient
  (exact or approximate) solutions to the robust design problems \cite{Gershman}.

Current robust design schemes can be classified into the worst-case and stochastic
approaches.
In the worst-case analysis \cite{Payaro}\cite{ShenoudaWorst}\cite{Vucic20092}\cite{Wang2}
\cite{Tajer}, the channel uncertainty is considered as deterministic and norm bounded.
The worst-case-based optimization approaches provide
 robustness against CSI imperfections.
However,
the actual worst case may occur  with a very low
probability. Hence, the worst-case approach may be overly pessimistic
and  therefore, may lead to unnecessary performance
degradation.
The
resulting optimization problem sometimes does not even have a feasible solution.
Even if the problem is feasible, the
resource utilization is inefficient as most system resources must
be dedicated to provide guarantees for the worst-case scenarios.

To provide
more flexibility than the worst-case designs,  the outage-probability-constrained robust designs
have also been recently developed \cite{ZLuo} \cite{Shenouda} \cite{VucicChance}\cite{YRong} \cite{Chalise} \cite{VNtranos}\cite{GZheng}\cite{MDing}\cite{XZhang}
where the convex optimization tools are
 employed to solve the resulting stochastic optimization  (also referred to as probabilistic programming) problems.
In these less conservative approaches the channel state and channel uncertainty
are considered as random processes.
Compared to the   worst-case approach, the stochastic approach
achieves better average performance while keeping the probability of the worst performance
low.
Unfortunately, probability-constrained stochastic optimization are known to be computationally
intractable except for a few special cases. In general,
such optimizations are difficult to solve as their feasible sets are
often nonconvex. In fact, finding feasible solutions to a generic
probability-constrained program is itself a challenging research
problem in the operations research community.

In this paper, we consider several power allocation problems in multiuser
MISO channels with imperfect CSI. We assume that the multi-antenna transmitters
employ some fixed beamformers to transmit the data and we focus on optimizing the
transmit powers of different users. In particular, for MISO interference
channels,
we treat two closely related
optimization problems: one is to minimize the total
transmit power subject to the SINR outage constraints, and the other is to maximize the achievable SINR
margin under the power constraint.
For MISO broadcast channels, we treat the MSE-constrained power minimization problem.
All these problems are formulated as probability-constrained
stochastic optimization problems.
The major
challenge in the stochastic optimization based method is to replace the probabilistic constraints with   deterministic
ones. We make use of the {\em Bernstein approximation} technique,  which is a recent advance in the field of probability-constrained programming \cite{Bernstein}, to transform the probabilistic constraints into deterministic
constraints that are conservative. The stochastic optimization problems are then transformed into convex optimization
problems.

The remainder of this paper is organized as follows.
In Section II, we formulate
 the Probability-constrained power optimization problems in MISO interference channels.
In Section III,  we propose solutions to the stochastic optimization problems based on the Berstein approximation
technique and the long-step logarithmic barrier cutting plane (LLBCP) algorithm.
In Section IV, we treat
  the MSE-constrained stochastic power optimization problem for  MISO broadcast channels.
Section IV presents the simulation results and finally,  conclusions are drawn in Section V.

\section{Power Control for  MISO Interference Channels with Imperfect CSI}

\subsection{System Model}

We consider a MISO interference channel with \emph{K} transmitters and
\emph{K} receivers.
Each transmitter employs $M$  transmit antennas and each
receiver is   equipped with a single receive antenna.
We assume that all receivers treat co-channel
interference as noise, i.e., they make no attempt to decode
the interference. Assuming a narrowband channel
model, the received signal at receiver $k$ is given by
\begin{equation}\label{Equ1}
y_{k}=\utwi{h}_{kk}^{H}\utwi{x}_{k}+\sum_{j\neq k}
\utwi{h}_{kj}^{H}\utwi{x}_{j}+n_{k}
 \end{equation}
where $\utwi{x}_{k} \in {\mathbb C}^M$ is the transmitted signal vector by the $k$-th transmitter,
$\utwi{h}_{kj} \in {\mathbb C}^M$ is the vector of complex channel gains between
the \emph{j}-th transmitter and the \emph{k}-th receiver,
$n_k \sim {\cal N}_c(0, \eta_{k}^{2})$ is the complex Gaussian noise sample.

We assume that each transmitter employs the beamforming technique to transmit information; that is,
we have $\utwi{x}_{k}=\utwi{w}_{k} s_{k}$, where $\utwi{w}_{k}\in {\mathbb C}^M$ is the
transmit beamformer for the link between the \emph{k}-th transmitter and the \emph{k}-th receiver, and $s_{k}
\in {\mathbb C}$ denotes the  complex
 data symbol intended for  the \emph{k}-th receiver.

In practice, the channel state information (CSI) at the receiver or transmitter is imperfect,
especially for the transmitter-side CSI.  In this paper,
 we assume that the transmitter only
has  access to the imperfect CSI to form the beamforming vectors. Specifically,
we have the following uncertainty model for the channel vectors
\begin{equation}\label{Equ2}
\utwi{h}_{kj} = \utwi{\hat {h}}_{kj}  +  {\utwi\delta} _{kj}
 \end{equation}
where $\utwi{\hat{h}}_{kj} = \left[ {\hat{h}_{kj}^1,...,\hat{h}_{kj}^M} \right]^T$ denotes
the imperfect  estimate of the actual channel vector $\utwi{h}_{kj}$,
and $\utwi{\delta} _{kj} = \left[ {\delta _{kj}^1,...,\delta _{kj}^M} \right]^T$ denotes the channel error
vector, where
\begin{equation}\label{delta}
\delta _{kj}^m \stackrel{\mbox{\small {i.i.d.}}}{\sim} \mathcal{N}_c\left( 0, \sigma _{kj}^2\right).
 \end{equation}
In order to obtain robust solutions that are less sensitive to  channel
uncertainties, we need to  explicitly take into account the impefect CSI. However, the mathematical problem arising from the robust
beamformer design  is in general much more
complicated than the conventional non-robust design. Thus some simplifications are needed
to make the problem tractable.
In this paper, following \cite{VucicChance,Payaro},
we assume that the transmit beamformer is of the form
 $\utwi{w}_{k}=p_{k}\utwi{g}_{k}$, where $p_{k} \in {\mathbb R}^+$ denotes transmit power, and
$\utwi{g}_{k}=[g_k^1,...g_k^M]^T \in {\mathbb{C}^{M}}$ denotes the unit-norm beamforming vector, i.e., $||\utwi{g}_k||_{2}=1$.
Here $\utwi{g}_{k}$  depends only on
the estimated channels and it is designed in a non-robust way.
For example, in the channel-matching approach, we have $\utwi{g}_{k} = \utwi{\hat{h}}_{kk}$.
On the other
hand, the design of the power allocation $p_{k}$ is much more
sophisticated and it depends not only on the channel
estimates, but also on the channel error statistic.
In this paper, we assume that the beam directions $\{ \utwi{g}_{k} \}$ are fixed and focus
 on the power allocation problem.

\subsection{Problem Formulations}
We next formulate the robust
power allocation problem based on the outage probability.
Our goal is to optimize the system performance through
power allocation. The system performance is usually
quantified by its quality of service (QoS) and the resources it
uses.
The most common QoS metrics are the symbol error rate or the achievable
data rate, both of which are functions of  the output SINRs.
The  system resources include the transmit power and bandwidth.
We consider two probability-constrained optimization problems as follows.

\subsubsection{SINR-constrained power minimization}

The first strategy seeks to minimize the average transmit
power subject to the QoS constraints.
Given an acceptable SINR level $\alpha_{k}$ and the outage probability
$\varepsilon_{k}$ for the \emph{k}-th  transmitter and receiver pair,
we aim to minimize the average transmit power while meeting the SINR
outage constraints of all users, i.e.,
\begin{eqnarray}\label{MinPower}
  \mathop {\min }\limits_{\left\{ {{p_k}} \right\}_{k = 1}^K} && \sum\limits_{k = 1}^K {{p_k}} \nonumber \\
  {\rm s.t.}\ \ &&\Pr \left( {{\Gamma _k} \leqslant {\alpha _k}} \right) \leqslant {\varepsilon _k},\ \ k =1, ..., K \\
 &&{p_k} \geqslant 0, \ k=1,...,K \nonumber
\end{eqnarray}
where $\mathrm{Pr}(A)$ denotes the probability of the event $A$, and $ {\Gamma _k} $ denotes the  SINR  at the \emph{k}-th receiver, given by
\begin{equation*}
    {\Gamma _k} = \frac{{{p_k}{{\left| {{\utwi{h}_{kk}^H}{\utwi{g}_k}} \right|}^2}}}
{{\eta _k^2 + \sum\limits_{j \ne k} {{p_j}{{\left| {{\utwi{h}_{kj}^H}{\utwi{g}_j}} \right|}^2}} }}
\end{equation*}
The design parameters $\varepsilon _k$  ensures that  receiver $k$ is served with an SINR no less than $\alpha _k$ at least $(1-\varepsilon _k) \times 100\%$ of the time.

\subsubsection{Max-Min SINR optimization}
The second strategy is to maximize the minimum SINR among all receivers, subject
to the SINR outage constraints, and the individual transmit power constraints.
We have the following optimization problem.
\begin{eqnarray}\label{MaxMinRate}
  \mathop {\max \min }\limits_{\left\{ {{p_k, \alpha_k}} \right\}_{k = 1}^K}&& {\alpha_k} \nonumber\\
  {\rm s.t.} &&\Pr \left( {{\Gamma _k} \leqslant \alpha_k} \right) \leqslant {\varepsilon _k},\ k=1, ..., K  \\
  &&0 \leqslant {p_k} \leqslant {{\bar p}_k},\ k=1, ..., K \nonumber
\end{eqnarray}
where ${\bar p}_k$ is the given power constraint for the \emph{k}-th transmitter.

Note that the problem in (\ref{MaxMinRate}) involves the  individual power constraint.
Alternatively we can also consider the total power constraint to have the following optimization problem.
\begin{eqnarray}\label{MaxMinRatetotal}
  \mathop {\max \min }\limits_{\left\{ {{p_k, \alpha_k}} \right\}_{k = 1}^K}&&  {\alpha_k} \nonumber \\
  {\rm s.t.} &&\Pr \left( {{\Gamma _k} \leqslant \alpha_k }\right) \leqslant {\varepsilon _k},\ k=1, ..., K \nonumber \\
 &&{p_k} \geqslant 0, \ k=1,...,K   \\
  &&\sum_{k=1}^K{p_k} \leqslant {{\bar p}_{tot}} \nonumber
\end{eqnarray}
where ${\bar p}_{tot}$ is the maximum allowable total transmit power.

\subsection{Background on Bernstein Approximation}

In problems (\ref{MinPower}), (\ref{MaxMinRate}), and (\ref{MaxMinRatetotal}),
the  probabilistic constraints  make the optimization
highly intractable. The main reason is that the convexity of the
feasible set corresponding to the  probabilistic constraints is difficult to verify.
To circumvent the above hurdles,
we make use of  the Bernstein approximation technique \cite{Bernstein} to convert
 the probabilistic constraints to convex constraints.
Next we briefly introduce the Bernstein approximation.

Suppose that $F ( \utwi{z}, \utwi{\zeta}): \mathbb{R}^{n} \times \mathbb{R}^{d}  \rightarrow \mathbb{R}$ is a function of $\utwi{z} \in \mathbb{R}^{n} $ and $\utwi{\zeta} \in \mathbb{R}^{d}$.
Then the probabilistic constraint
\begin{equation}\label{Ber3B}
\mathrm{Pr}\{F(\utwi{z}, \utwi{\zeta}) \geq 0\} \leq\ \varrho
\end{equation}
can be conservatively approximated by the following
\begin{eqnarray}\label{Ber3}
&& \mathrm{inf}_{t>0}\;\; \{ t {\mathbb E}\{ \psi [ t^{-1} F(\utwi{z}, \utwi{\zeta})] \} -t \varrho\} \leq 0,
\end{eqnarray}
or
\begin{eqnarray}\label{Ber3A}
&& \mathrm{inf}_{t>0}\;\; \{ t \log{\mathbb E}\{ \psi [ t^{-1} F(\utwi{z}, \utwi{\zeta})] \} -t  \log \varrho\} \leq 0.
\end{eqnarray}
where $ \psi: \mathbb{R}  \rightarrow \mathbb{R}$
be a nonnegative valued, nondecreasing, convex function satisfying
$ \psi(z) > \psi(0) = 1$ for any $z > 0$.
For example, $ \psi(z)=\exp(z)$.
Moreover,
assume that for every $\utwi{\zeta} \in \mathbb{R}^{d}$ the function $F(\cdot, \utwi{\zeta})$ is convex. Then
$t {\mathbb E}\{ \psi [ t^{-1} F(\utwi{z}, \utwi{\zeta})] \} -t \varrho$ is convex, and thus (\ref{Ber3}) is convex.
Indeed, since $ \psi(\cdot)$is nondecreasing and convex and $F(\cdot, \utwi{\zeta})$is convex, it
follows that $(z, t) \rightarrow t\psi(t^{-1}F(z, \utwi{\zeta}))$ is convex. This, in turn, implies convexity of the
expected value function ${\mathbb E}\{ \psi [ t^{-1} F(\utwi{z}, \utwi{\zeta})] \}$, and hence convexity of $t {\mathbb E}\{ \psi [ t^{-1} F(\utwi{z}, \utwi{\zeta})] \} -t \varrho$. Similarly, $t \log {\mathbb E}\{ \psi [ t^{-1} F(\utwi{z}, \utwi{\zeta})] \} -t \log \varrho$ is convex, and thus (\ref{Ber3A}) is convex.

As an important special case,
suppose that $\utwi{\zeta}$  is a random vector whose components are independent and nonnegative.
$F (\utwi{z}, \utwi{\zeta})$ is affine in $\utwi{\zeta}$, i.e.,
\begin{equation}\label{BerTheorem1}
    F( \utwi{z}, \utwi{\zeta})=f_{0}(\utwi{z})+\sum_{j=1}^{d}\zeta_{j} f_{j}(\utwi{z}),
\end{equation}
and the functions $f_{j}(\utwi{z}),  j =0, 1,...,d$, are well defined and convex on $\mathbb{R}^{n}$.
Then the probabilistic constraint (\ref{Ber3B}) can be conservatively approximated by (\ref{Ber3W}).
\begin{equation}\label{Ber3W}
 \mathrm{inf}_{t>0}\;\; \{ t \log {\mathbb E}\{ \exp [ t^{-1} F(\utwi{z}, \utwi{\zeta})] \} -t\log \varrho\} 
 = \mathrm{inf}_{t>0}\;\; \{ f_{0}(\utwi{z})+\sum_{j=1}^{d}t \log \mathbb{E}\{\exp[t^{-1}\zeta_{j}f_{j}(\utwi{z})]\} -t\log \varrho\}\leq 0.
\end{equation}
Furthermore, the constraint  in (\ref{Ber3W}) is  convex.
To see this, define
\begin{eqnarray}\label{BerTheorem2}
\begin{gathered}
\Psi(\utwi{\gamma})
\triangleq  \log (\mathbb{E} [\exp\{\gamma_{0}+\sum_{j=1}^{d}\zeta_{j}\gamma_{j} \} ] ) \hfill \\
\;\;\;\;\;\;\;\;\;=\gamma_{0}+\sum_{j=1}^{d}\log\mathbb{E}[\exp(\zeta_{j}\gamma_{j})].\hfill \\
\end{gathered}
\end{eqnarray}
Then (\ref{Ber3}) can be written as
\begin{eqnarray}
\mathrm{inf}_{t>0}\;\;  \{ t\Psi(t^{-1} \utwi{f}(\utwi{z})) - t \log\varrho \} \leq 0.
\end{eqnarray}
The function
$G(\utwi{\gamma}, t) \triangleq t\Psi(t^{-1}\utwi{\gamma})-t\log\varrho$
is convex in $(\utwi{\gamma}, t > 0)$ (since $\Psi(\utwi{\gamma})$ is convex) and is nondecreasing in $\gamma_{0}$ and every $\gamma_{j}, j=1,2,...,d$ (since $\zeta_j>0$).
Since all $f_{j}(\utwi{z}), j = 0,1,...,d$ are convex, then  $G(\utwi{f}(\utwi{z}), t)$ is
convex. Due to preservation of convexity by minimization over  $t > 0$, (\ref{Ber3W})  is convex.

Note that  by using the Bernstein approximation, we can convert the
intractable optimization problem with probabilistic constraint into
an explicit convex optimization.

\section{Robust Power Optimization for MISO Interference Channels}

In this section, we apply the Bernstein approximation to obtain the
convex approximations to  the probabilistic constraints
in  problems (\ref{MinPower}), (\ref{MaxMinRate}), and (\ref{MaxMinRatetotal}),
and then solve the resulting convex problems using the long-step logarithmic barrier
cutting plan (LLBCP) algorithm.

\subsection{Robust SINR-constrained Power Minimization}

The major difficulty in the robust power optimization design is to convert the probabilistic constraint into a deterministic one.
To that end we apply the Bernstein approximation to obtain
the following result. The proof is given in Appendix A.

\noindent \textbf{Proposition 1}
The following optimization problem (\ref{problem2}) is a convex conservative approximation to the optimization problem in (\ref{MinPower}):
\begin{eqnarray}\label{problem2}
  \mathop {\min }\limits_{\left\{ {{p_k}}\right\}_{k = 1}^K}&&\sum\limits_{k = 1}^K {{p_k}}\nonumber\\
 {\rm  s.t.} && \mathop {\inf }\limits_{{t_k} >\rho_k }  {G_k}\left( {{\utwi{p}},{t_k}} \right) \leqslant 0, \ k=1,...,K \label{Feasible1}  \notag  \\
  &&{p_k} \geqslant 0, \ k=1,...,K
\end{eqnarray}
where ${G_k}\left( {{\utwi{p}},{t_k}} \right)$ is defined in (\ref{Gk1})

 \begin{eqnarray}\label{Gk1}
{\text{      }}{G_k}\left( {{\utwi{p}},{t_k}} \right)
&\triangleq& 
 {\alpha _k}\eta _k^2
   + {t_k}\sum\limits_{j \ne k} { {\left[ {\frac{{t_k^{ - 1}{\alpha _k}{p_j}{{\left| {\utwi{\hat {h}}_{kj}^H \utwi{g}_j } \right|}^2}}}
{{1 - t_k^{ - 1}{\alpha _k}{p_j} {\sigma_{kj}^2} {{\left| {\boldsymbol{1}^T \utwi{g}_j } \right|}^2}}} - \log \left( {1 - t_k^{ - 1}{\alpha _k}{p_j}\sigma _{kj}^2{{\left| {\boldsymbol{1}^T \utwi{g}_j } \right|}^2}} \right)} \right]} }  \hfill \\ \notag
   &&- { {\frac{{{p_k}{{\left| {\utwi{\hat {h}}_{kk}^H \utwi{g}_k } \right|}^2}}}
{{1 +  t_k^{ - 1}{p_k}\sigma_{kk}^2{{\left| {\boldsymbol{1}^T \utwi{g}_k } \right|}^2}}} - {t_k} \log \left( {1 +  t_k^{ - 1}{p_k}
\sigma_{kk}^2{{\left| {\boldsymbol{1}^T \utwi{g}_k} \right|}^2}} \right)} }  - {t_k}\log \left( {{\varepsilon _k}} \right)
\end{eqnarray}
, and \begin{eqnarray}
 \ \ \  {\rho_k} &\triangleq&  \alpha_k \max_{ j \neq k}  \{   \sigma^2_{kj}p_j | {\boldsymbol{1}^T \utwi{g}_j}  |^2   \}
\end{eqnarray}

\subsection{Long-step Logarithmic Barrier Cutting Plane (LLBCP) Algorithm  }\label{LLBCPA}
Notice that the first constraint in  (\ref{problem2})
  is itself in terms of an optimization problem. Hence, although the optimization problem  (\ref{problem2}) is
convex, it cannot be straightforwardly  solved
using a standard convex optimization solver.
We will employ the long-step logarithmic barrier cutting plane  (LLBCP) algorithm  to solve it.

The detailed development of  the LLBCP algorithm  is found in \cite{Mitchell2000}\cite{Mitchell}.
Here we outline the basic ideas of this method.
Suppose that we would like to find a solution $\utwi{p}$ that is feasible for
(\ref{problem2}) and satisfies $\|\utwi{p}-\utwi{p}^{*}\|<\epsilon$\footnote{$\|\cdot\|$ denotes Euclidean norm operator.} for some optimal solution
$\utwi{p}^{*}$ to (\ref{problem2}), where $\epsilon > 0$ is the error tolerance parameter\footnote{
It is assumed that there exist the set of feasible solutions to (\ref{problem2}),
and a problem dependent constant $\epsilon$ such that

\begin{enumerate}
  \item The set of optimal solutions to (\ref{problem2}) is guaranteed to be contained in the $K$ dimensional
   hypercube of half-width $1/\epsilon$.
  \item The set of feasible solutions to (\ref{problem2}) contains a full dimensional ball of radius $\epsilon$.
  \item It suffices to find a solution to (\ref{problem2}) to within an accuracy $\epsilon$.
\end{enumerate}}.
At the beginning of each iteration, the feasible
set, if exists, is contained in a bounded polytope. Then, we generate a trial point by constructing the analytic center inside the bounded polytope, and test  whether or not the trial point belongs to the feasible set.
If this trial point is not feasible,
a hyperplane through the trial point is introduced to
cut off the violated constraint(s), so that the remaining polytope contains
the feasible set.
When the trial point is feasible but not optimal,  by
updating the lower bound on the optimal objective function value of  problem (\ref{problem2}) and reducing the barrier parameter, the new optimality constraint(s) is generated to update the polytope.
Furthermore, if the hyperplanes currently in the polytope are
deemed ``unimportant'' according to some criteria, they are dropped.
We can then proceed to the next iteration with the new
polytope until the termination condition is satisfied.

Assuming  that there exist the set of feasible solutions to (\ref{problem2}),
as shown in \cite{Mitchell2000}\cite{Mitchell}, there are three termination conditions in the LLBCP algorithm:
\begin{enumerate}
  \item \emph{\textit{Termination 1:}}  The number
of hyperplanes exceeds a certain level, so that the volume of the current polytope would be too small to contain a small
enough ball.
  \item \emph{\textit{Termination 2:}}  The smallest slack is smaller than
a certain number, so that the polytope would be
too narrow to contain a small enough ball.
  \item \emph{\textit{Termination 3:}}  The duality gap
is enough small, so that the algorithm may be terminated with optimality.
\end{enumerate}

In terms of convergence, as shown in \cite{Mitchell2000}\cite{Mitchell},
the LLBCP algorithm  terminates with a solution $\utwi{p}$ that is feasible for
(\ref{problem2}) and satisfies $\|\utwi{p}-\utwi{p}^{*}\|<\epsilon$ for some optimal solution
$\utwi{p}^{*}$ to (\ref{problem2}) after at most $\mathcal{O}(K (\log_{2}(1/\epsilon ))^{2})$ iterations, where $K$ is the number of variables.
Note that although the LLBCP algorithm  has the same order of complexity as the  algorithm in \cite{Atkinson},\cite{SlowAdaptive},
in practice it  is computationally much more efficient \cite{Mitchell2000}.

\begin{figure*}[htbp]
 \centering
 \includegraphics[scale=1,bb=37 498 529 815]{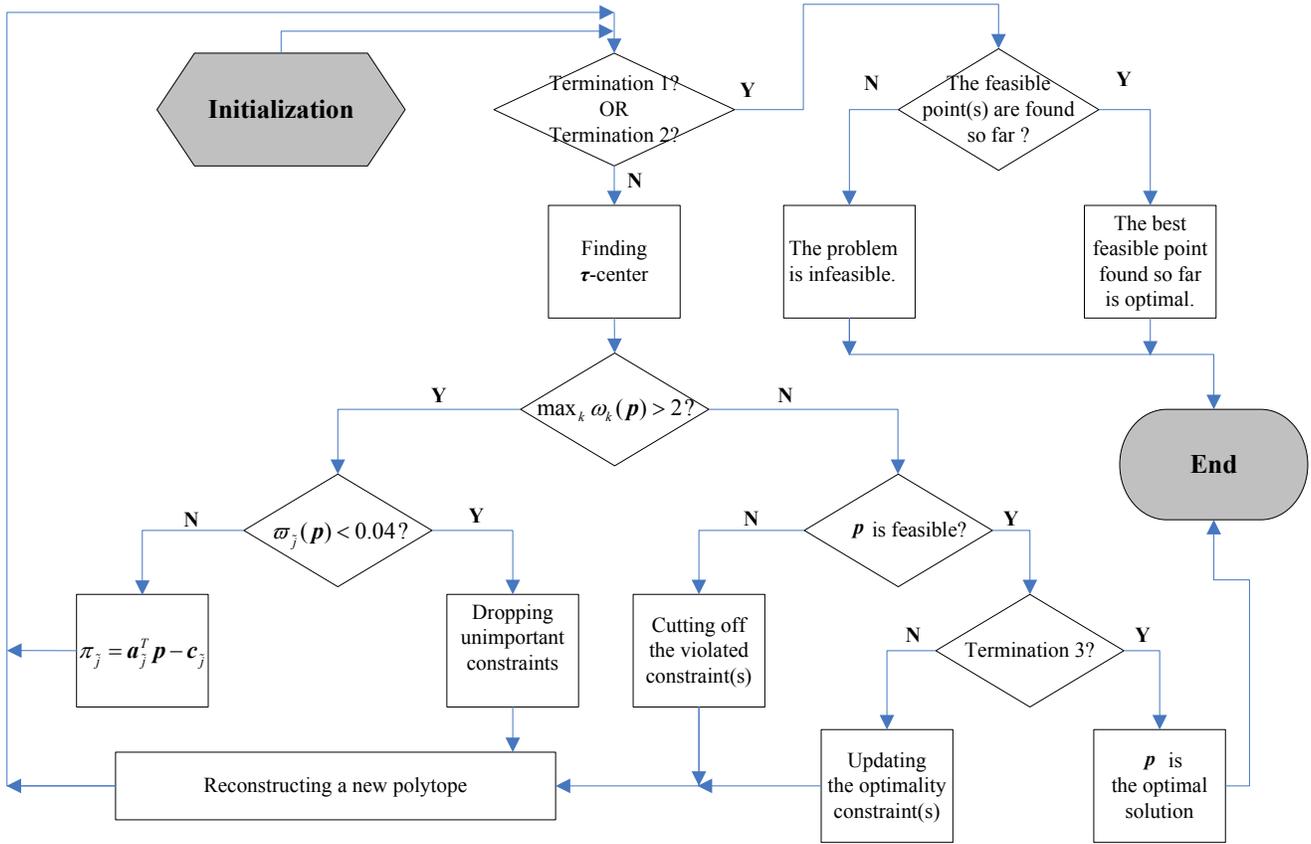}
 \caption{Flow chart of the  LLBCP algorithm. }
 \label{FlowChart}
 \end{figure*}

In Fig.~\ref{FlowChart} we give a
 detailed flow chart of the LLBCP algorithm; and in Algorithm 1, we  give
 the step-by-step description of the algorithm. The key components
 of the LLBCP algorithm are then elaborated next.


\baselineskip 18pt

\begin{algorithm}
\caption{: LLBCP Algorithm}
\begin{algorithmic}[1]

\State Initialization
\State  Set $\epsilon > 0$, $\tau = \frac{1}{\epsilon}$, $i = 0$;
\State According to (\ref{problem2}), set

   $\utwi{A}=[\utwi{I}_{K} \;\;-\utwi{I}_{K}\;\;\utwi{1}_{K}]^{T}$,
   $\utwi{c}=[-\frac{1}{\epsilon}\utwi{1}_{K}^{T}\;\;-\frac{1}{\epsilon}\utwi{1}_{K}^{T} \;\;-\frac{1}{\epsilon}\sqrt{K}]^{T}$,
$\utwi{p}=\utwi{0}_{K}$\footnote{where $\utwi{I}_{K}$, $\utwi{1}_{K}$  and $\utwi{0}_{K}$ denoting the $K \times K$ identity matrix, $K$ vector of ones, and $K$ vector of zeros, respectively.},
and $\utwi{s}=\utwi{A}\utwi{p}-\utwi{c}$;
\State  Set $\pi_{k}=\frac{1}{\epsilon}, k=1,...,2K,\pi_{2K+1}=\frac{1}{\epsilon}\sqrt{K};$
\State  If necessary, find approximate $\tau$-center.

\State  The Iterative step

\If  {\emph{\textbf{Termination 1}}: $N \geq 4093K\log_{2}(1/\epsilon )$ \;\;\;\;\;\;\;\;\;\;\;\;\;\;\;\;\;\;\;\;\;\;\;\;\;\;\;\;\;\;\;\;\;\;\;\;\;\;\;\;\;\;\;\;\;\;\;\;\;\;\;\;\;\;\;\;\;\;\;\;\;\;\;\;\;\;\;\;\;\;\;\;\;\;\;\;\;\;\;\;\;\;\;\;\;\;\;\;\;\;\;\;\;\;\;\;\;\;\;\;\;\;\;\;\;\;\;\;\;\;\;\;\;\;\;\;\;\; \;\;\; or \emph{\textbf{Termination 2}}: $\min_{k}({\utwi{a}^{T}_{k}\utwi{p} - c_{k}}) < \;\;\; 10^{-5}\epsilon^{3}/[2K^{1.5}\log_{2}(1/\epsilon )]$}

\State STOP: the best feasible point found so far is optimal. Otherwise, no feasible point is found.

\Else
\State Find a new approximate $\tau$-center $\utwi{p}=\utwi{p}^{i}$;
 \State Calculate $\omega_{k}(\utwi{p}) = \frac{\utwi{a}^{T}_{k}\utwi{p}- c_{k}}{\pi_{k}}, \forall k$.
\State  Set $\omega_{k}(\utwi{p})=1$  if $k$ indexes the lowerbound
constraint that get added in \textbf{Subcase 2.2}.
\State  If $\omega_{\tilde{j}}(\utwi{p})>2$, calculate $\varpi_{\tilde{j}}(\utwi{p})$ as in (\ref{variational quantities}).

\If {  $\max_{k}(\omega_{k}(\utwi{p}))>2$}
\State  \textbf{Case 1} :

 \If {for some $\tilde{j}$, we have $\varpi_{\tilde{j}}(\utwi{p})<0.04$}
\State   \textbf{Subcase 1.1}:
\State  Drop the hyperplane $\utwi{a}_{\tilde{j}}$;
\State  Find a new approximate $\tau$-center.
\Else
\State  \textbf{Subcase 1.2}:
\State Reset $\pi_{\tilde{j}} = \utwi{a}^{T}_{\tilde{j}}\utwi{p} - c_{\tilde{j}}$, where $\tilde{j} $ be an index such that $\omega_{\tilde{j}}(\utwi{p})>2$.
\EndIf
\EndIf

\If { $\max_{k}(\omega_{k}(\utwi{p}))\leq 2$}
\State  \textbf{Case 2} :
\If {$\utwi{p}$ is not feasible in the problem  (\ref{problem2})}

\State \textbf{Subcase 2.1}:
\State For $\tilde{k} \in \tilde{\utwi{K}}$,
generate hyperplane(s) as in (\ref{ViolatedCut}).
\State  Set $\pi_{N+\tilde{k}}=(\frac{\utwi{\nabla}_{\mathrm{drop}}^{i,\tilde{k}}}{\|\utwi{\nabla}_{\mathrm{drop}}^{i,\tilde{k}}\|})^{T}\utwi{p}^{i}- (\frac{\utwi{\nabla}_{\mathrm{drop}}^{i,\tilde{k}}}{\|\utwi{\nabla}_{\mathrm{drop}}^{i,\tilde{k}}\|})^{T}\utwi{p}$.
\State  Set $N \leftarrow N + |\tilde{\utwi{K}}|$\footnote{$|\cdot|$ is the
number of elements contained in the given set}.
\State  Find a new approximate $\tau$-center\footnote{If no more than $ |\tilde{\utwi{K}}|$ cuts are added simultaneously, the total number of Newton steps is multiplied by a factor of at most $O( |\tilde{\utwi{K}}| \log( |\tilde{\utwi{K}}|))$\cite{Goffin}.}.

\Else
\State \textbf{Subcase 2.2}:

\If  {\emph{\textbf{Termination 3}}: $1.25N\tau < \epsilon$}
\State  $\utwi{p}_{i}$ is the optimal solution, and STOP.
\Else
 \State  set the lower bound $l = \utwi{1}_{K}^{T}\utwi{p}- 1.25N\tau$ on optimal objective function of (\ref{problem2}).
\State   Let $l_{prev}$ denote previous lower bound. If $l_{prev} < l$, replace $\utwi{1}_{K}^{T}\utwi{p}\geq l_{prev} $ by $\utwi{1}_{K}^{T}\utwi{p}\geq l$.
\State   Set $\tau \leftarrow \theta\tau$, where $\theta \in (0.5, 1)$.
\State   Find a new approximate $\tau$-center.
\EndIf
\EndIf
\EndIf
\EndIf
\end{algorithmic}
\end{algorithm}

\subsubsection{Finding $\tau$-center}(Lines 5, 10, 19, 32, 41 in Algorithm 1)

In each iteration \emph{i}, we need to generate a  trial point inside the polytope $\mathcal{{P}}^{i}=\{\utwi{p}\in \mathbb{R}^{K}:{\utwi{A}^{i}\utwi{p}\geq \utwi{c}^{i}}\}$.
Here, we will generate the so-called $\tau$-center of the  polytope $\mathcal{{P}}^{i}$ as the trial point.
First we need to define the so-called logarithmic barrier function
\begin{equation}\label{BarrierFunction}
 f(\utwi{p},\tau)=\frac{\utwi{1}_K^{T}\utwi{p}}{\tau}-\sum_{n}\log(s_{n})
 \end{equation}
where $s_{n}=\utwi{a}_{n}^{T}\utwi{p}-c_{n}$, and $\utwi{a}_{n}^{T}$
is the \emph{n}-th row of $\utwi{A}$.
$\tau> 0$ is the barrier parameter.
For a given value of $\tau^{i}$, $\utwi{p}^{i} (\tau^{i})$ denotes the unique minimizer of this barrier function.
We refer this unique point as the $\tau$-center.
Notice that an approximate
$\tau$-center is sufficient to serve as a  trial point. An approximate $\tau$-center
for the $(i+1)$-th iteration can be obtained from an approximate
$\tau$-center for the $i$-th iteration by applying $\mathcal{O}(1)$ Newton steps\cite{Goffin}.

\subsubsection{Dropping unimportant constraints}(Lines 14-20 in Algorithm 1)


The $j$-th constraint (i.e., hyperplane) is dropped only if its slack $\utwi{a}_{j}^{T}\utwi{p}- c_{j}$ has doubled since $\omega_ j$ was last reset (Line 14 in Algorithm 1) and its variational quantity $\varpi_j$ is small (Line 16 in Algorithm 1).

The variational quantity $\varpi_j$ is defined as:
\begin{equation}\label{variational quantities}
 \varpi_j=\frac{\utwi{a}_{j}^{T} (\nabla^{2} f(\utwi{p},\tau))^{-1}  \utwi{a}_{j}}{s_{j}^{2}},\;\; j=1,...,N,
\end{equation}
where $N$ denote the number of the rows of $\utwi{A}$.
The variational quantities give an indication of the relative importance
of the constraint $\utwi{a}_{j}^{T}\utwi{p}\geq c_{j}$.

If $k$ indexes the lower bound, $\omega_{k}(\utwi{p})=1$ (Line 12 in Algorithm 1). Otherwise,  $\omega_{k}(\utwi{p}) = \frac{\utwi{a}^{T}_{k}\utwi{p}- c_{k}}{\pi_{k}}$ (Line 11 in Algorithm 1), where $\pi_{k}$ is initialized as $\pi_{k}=\frac{1}{\epsilon}, k=1,...,2K,\pi_{2K+1}=\frac{1}{\epsilon}\sqrt{K}$ (Line 4 in Algorithm 1), and is updated as $\pi_{k}=\utwi{a}^{T}_{k}\utwi{p} - c_{k}$ (Line 22 in Algorithm 1) if
$\max_{k}(\omega_{k}(\utwi{p}))>2$(Line 14 in Algorithm 1) and $\varpi_{k}(\utwi{p}) \geqslant 0.04$ (Line 16 in Algorithm 1).

\subsubsection{Checking feasibility}

Given a trial point $\utwi{p}^{i} \in \mathcal{{P}}^{i}$, we can verify its feasibility for problem  (\ref{problem2}) by checking if it satisfies the first constraint $\mathop {\inf }\limits_{{t_k} >\rho_k }  {G_k}\left( {{\utwi{p}}^{i},{t_k}} \right) \leqslant 0, \forall k$.
This requires solving a minimization problem over  $t_k >\rho_k$.
Due to the unimodality of ${G_k}\left( {{\utwi{p}}^{i},{t_k}} \right)$ in $t_k $,
we can simply take a line search procedure to find the minimizer $t_k^{*} $.

\subsubsection{Cutting off the violated constraint(s)} (Lines 29-31 in Algorithm 1)

If the trial point $\utwi{p}^{i} \in \mathcal{{P}}^{i}$ is infeasible, then a hyperplane
is generated at  $\utwi{p}^{i}$ as follows:
\begin{equation}\label{ViolatedCut}
 \Big(\frac{\utwi{\nabla}_{\mathrm{drop}}^{i,\tilde{k}}}{\|\utwi{\nabla}_{\mathrm{drop}}^{i,\tilde{k}}\|}\Big)^{T}\utwi{p}\leq \Big(\frac{\utwi{\nabla}_{\mathrm{drop}}^{i,\tilde{k}}}{\|\utwi{\nabla}_{\mathrm{drop}}^{i,\tilde{k}}\|}\Big)^{T}\utwi{p}^{i}, \;\;\; \tilde{k} \in \tilde{\utwi{K}}
\end{equation}
where 
$\tilde{\utwi{K}}=\{k:{G_k}\left( {{\utwi{p}}^{i},{t_k^{*}}} \right) > 0, \; k=1,...,K\}$, and
$\utwi{\nabla}_{\mathrm{drop}}^{i,\tilde{k}}$ is the gradient
of ${G_k}\left( {{\utwi{p}},{t_k^{*}}} \right)$ with respect to $\utwi{p}$ at $\utwi{p}^{i}$,
with the \emph{k}-th component given by
\begin{equation}
  \nabla_{\mathrm{drop},k}^{i,\tilde{k}}=\frac{\partial{G_{\tilde{k}}\left( {{\utwi{p}},{t_{\tilde{k}}^{*}}} \right)}}{\partial {p_k}}|_{p_k={p_{k}^{i}}}
\end{equation}

\subsubsection{Updating lower bound and reducing barrier parameter} (Lines 38-40 in Algorithm 1)
If the point  $\utwi{p}^{i}$ is feasible but not optimal, the lower bound $l = \utwi{1}_{K}^{T}\utwi{p}- 1.25N\tau$ to the optimal objective function value of   problem (\ref{problem2}) is updated (Lines 38-39 in Algorithm 1), and the value of the barrier parameter $\tau$ is  reduced (Line 40 in Algorithm 1).
Notice that according to the definition of (\ref{BarrierFunction}),
for a fixed value of $\tau > 0$, it is desirable to minimize $ f(\utwi{p},\tau)$, leading to a balance between the objective function and centrality.
When we need to drive the objective function value down,  we just reduce the value of the barrier parameter $\tau$, leading to increasing emphasis
on the objective function .
When $\tau$ is driven to zero,  we have the convergence to an optimal solution.

\subsection{Robust Max-Min SINR Optimization}

We next consider the max-min SINR optimization problem in
  (\ref{MaxMinRate}).  Since it is difficult to verify directly whether  problem (\ref{MaxMinRate}) is convex, we use the similar method in \cite{Tajer} to solve  (\ref{MaxMinRate}).  Specifically,
by introducing a slack variable $a > 0$, the epigraph form of the robust max-min SINR optimization
problem with individual power constraints (\ref{MaxMinRate}) is given by
\begin{equation}
    \mathcal{S}({\utwi{ p}}) \triangleq \left\{ \begin{gathered}
  \mathop {\max }\limits_{\left\{ {{p_k}} \right\}_{k = 1}^K,a} {\text{  }}a \hfill \\
 {\rm  s.t.}{\text{  }}\Pr \left( {{\Gamma _k} \leqslant a } \right) \leqslant {\varepsilon _k}, \ k=1,...,K \hfill \\
  {\text{       }}0 \leqslant {p_k} \leqslant {{\bar p}_k},\ k=1,...,K \hfill \\
\end{gathered}  \right.
\end{equation}
We demonstrate that solving $\mathcal{S}({\utwi{p}})$ can be facilitated via solving a power
optimization problem defined as
\begin{equation}
  \mathcal{P}({\utwi{p}},a)  \triangleq  \left\{ \begin{gathered}
  \mathop {\min }\limits_{\left\{ {{p_k}} \right\}_{k = 1}^K,b} {\text{ }}b \hfill \\
  {\rm s.t.}{\text{  }}\Pr \left( {{\Gamma _k} \leqslant a } \right) \leqslant {\varepsilon _k},\ k=1,...,K\hfill \\
  {\text{       }}0 \leqslant {p_k} \leqslant b{{\bar p}_k},\ k=1,...,K \hfill \\
\end{gathered}  \right.
\end{equation}
which can be solved using the similar method for solving the robust power minimization problem  given in (\ref{MinPower}).
The connection between $\mathcal{S}({\utwi{ p}})$ and $\mathcal{P}({\utwi{p}},a)$ is given by the following result, and the proof is given in Appendix B.

\noindent \textbf{Proposition 2}  $\mathcal{P}({\utwi{ p}},a)$ is strictly increasing and continuous in $a$ at any strictly feasible region and is
related to $\mathcal{S}({\utwi{p}})$ via $\mathcal{P}({\utwi{ p}},\mathcal{S}({\utwi{ p}})) =$1.

Since $\mathcal{P}({\utwi{ p}},a)$ is strictly increasing and continuous in $a$ at any strictly feasible region, there exists a unique $a^{*}$ satisfying $\mathcal{P}({\utwi{p}},a^{*})$ = 1. It follows from  Proposition 2 that solving
$\mathcal{S}({\utwi{p}})$ boils down to finding $a^{*}$ that satisfies $\mathcal{P}({\utwi{p}},a)=1$. Due to monotonicity and continuity of
$\mathcal{P}({\utwi{p}},a)=1$,   $a^{*}$ can be obtained by a simple bi-section search.

Finally we note that using the same approach, we can  solve the problem of robust max-min SINR optimization
problem with total power constraints given in (\ref{MaxMinRatetotal}).

\section{Robust Power Optimization for   MISO  Broadcast Channels}

In this section, we treat a related power allocation problem for a MISO broadcast system
with outage constraints on receiver MSE. Specifically,
 we consider the MSE-constrained power minimization problem in the downlink multiuser MISO system with Gaussian channel mismatch.
We adopt the Bernstein approximation approach to convert the probablistic constraint into a deterministic convex constraint.
Note that the similar problem has been considered in \cite{VucicChance}, where the Vysochanskii-Petunin inequality (VPI) is employed to  obtain the
conservative approximations to the probablistic constraints. We will demonstrate in Section IV the superiority of the Bernstein approach to the VPI method.

We consider a downlink multiuser MISO system with one base station (BS) equipped with $M$ antennas
and $K$ single-antenna users.
The BS transmits a symbol vector $\utwi{s}=[s_{1},\ldots,s_{K}]^T \in \mathbb{C}^K$, where
the symbol $s_{k}$ is intended for the $k$-th  user.
We denote
the complete downlink channel as $\utwi{H}\in \mathbb{C}^{K\times M}$. The BS is provided
only with an estimate $\hat{\utwi{H}}$ of $\utwi{H}$, where $\hat{\utwi{H}}$ has  full row
rank. The CSI error matrix is given by $\utwi{\Delta}=\utwi{H}-\hat{\utwi{H}}$, which is assumed
to contain i.i.d. complex Gaussian entries, i.e.,
\begin{eqnarray}
\delta_{k,j} \stackrel{\rm i.i.d.}{\sim}
{\cal N}_c(0, \sigma_{k,j}^2), k=1,..., K, j=1, ..., M. \label{ttt3.eq}
\end{eqnarray}

We assume that the beamforming matrix
$\utwi{G} \in \mathbb{C}^{M \times K}$
 is set as the Moore-Penrose
pseudoinverse of the available
imperfect CSI, i.e., $\utwi{G}=\hat{\utwi{H}^{\dag}}$.
Our objective is to design the diagonal power allocation matrix
$\utwi{Q}^{1/2}=\mathrm{diag}(\sqrt{q_{1}}, \ldots,\sqrt{q_{K}})$.

The $k$-th user equalizes its received
signal using a one-tap equalizer with coefficient $q_{k}^{-1/2}$. Thus the
symbol estimate at the equalizer output is given by
\begin{equation}
    \hat{s}_{k}=q_{k}^{-1/2}\utwi{H}[k,:]\utwi{G}\utwi{Q}^{1/2}\utwi{s}+v_{k},\ \ \ k=1,\ldots,K \label{ttt1.eq}
\end{equation}
where $\utwi{H}[k,:]$, denoting the
$k$-th row of $\utwi{H}$, is the $k$-th user's MISO channel;
$v_{k}$ denotes the noise sample at the $k$-th user.
We assume that $\mathbb{E}\{\utwi{ss}^{H}\}=\utwi{I}$, and
$\mathbb{E}\{\utwi{vv}^{H}\}=\mathrm{diag}(\eta_{1}^{2},\ldots,\eta_{K}^{2})$, where $\utwi{v}=[v_{1},\ldots ,v_{K}]^{T}$.

We  use
the mean-squared error (MSE) between the transmitted symbol and the receiver equalizer output
as the QoS metric, i.e.,
\begin{equation}
    \mathrm{MSE}_{k}=\mathbb{E}\{|s_{k}-\hat{s}_{k}|^{2}\},\ \ \ k=1,\ldots,K \label{ttt2.eq}
\end{equation}
We consider the following MSE-constrained power minimization problem. The objective is to minimize
the total transmit power, subject to  constraint that the probability of $\mathrm{MSE}_{k}$
being below a target value $\mu_k$ is no less than $\phi_{k}\in(0,1)$. That is,
\begin{eqnarray}\label{problemMSE}
\begin{gathered}
  \mathop {\min }\limits_{\utwi{Q}  \succeq {\bf 0}} \ \ \mathrm{tr}(\utwi{GQG}^{H})  \hfill \\
 {\rm  s.t.}
    \ \ \ \mathrm{Pr} \left(\mathrm{MSE}_{k}\leq \mu_k \right) \geq \phi_{k},\ k=1, ..., K \hfill \\
\end{gathered}
\end{eqnarray}
Now using the Bernstein approximation we can convert  (\ref{problemMSE})  into a
convex optimization problem, as stated in the following result.
The proof is given in Appendix C.

\noindent \textbf{Proposition 3}\
The following optimization problem (\ref{problemMSE_app}) is a convex conservative approximation to
the optimization problem in (\ref{problemMSE}):
\begin{eqnarray}\label{problemMSE_app}
\begin{gathered}
  \mathop {\min }\limits_{\utwi{Q} \succeq {\bf 0} } \ \ \mathrm{tr}(\utwi{GQG}^{H})  \hfill \\
  {\rm s.t.}
  \ \ \ \mathop {\inf }\limits_{{t_k} > 0 }  {G_k^{\mathrm{MSE}}}\left( {\utwi{Q},{t_k}} \right) \leqslant 0,\ k=1,...,K   \hfill \\
\end{gathered}
\end{eqnarray}
where $G_k^{\mathrm{MSE}} \left( {\utwi{Q}, {t}_k } \right)$ is defined in (\ref{GkMSE})

\begin{eqnarray}\label{GkMSE}
G_k^{\mathrm{MSE}} \left( {\utwi{Q}, {t}_k } \right) &\triangleq &
(\eta_k^2  - q_k \mu _k)
 - \frac{ {t}_k}{2} \log \det\Big(\utwi{I}-\frac{2}{ {t}_k}\mathbf{\Lambda}_k \utwi{GQG}^{H} \Big)-  t_k \log(1-\phi_k),
\end{eqnarray}
with $\utwi{\Lambda}_k
 \triangleq {\rm diag}( \sigma_{k,1}^2, ..., \sigma_{k,M}^2)$.

\medskip

We can solve   problem  (\ref{problemMSE_app}) using the LLBCP algorithm discussed in Section \ref{LLBCPA}.

\section{Simulation Results}
In this section, we present extensive simulation results to illustrate the performance of proposed Bernstein approximation approach to probability-constrained power optimization in wireless networks.
First we illustrate the   performance in  MISO interference channels.
Then we illustrate the performance in MISO broadcast channels and
compare it with that of the VPI-based approach given in \cite{VucicChance}.

\subsection {SINR-constrained Power Minimization in MISO Interference Channels}
\begin{figure}[htbp]
 \centering
 \includegraphics[scale=0.6,bb=21 275 490 513]{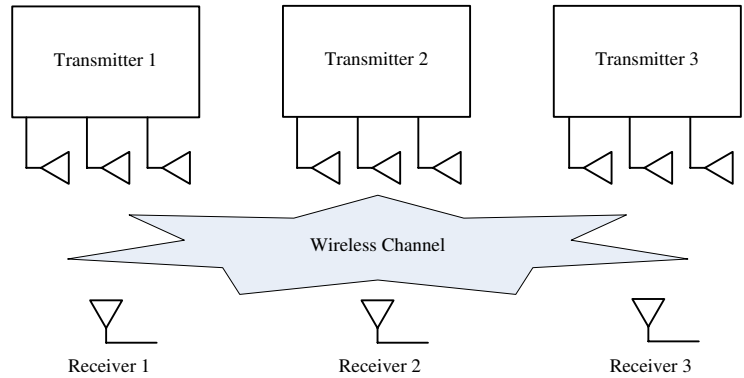}
 \caption{The simulated MISO interference channel.}
 \label{NetworkScenario}
 \end{figure}

We consider a MISO interference channel shown in Fig. \ref{NetworkScenario},
where the distance from  a
transmitter to the corresponding receiver is 200m, and  the distance between the adjacent
transmitters or receivers is 400m.
The channel from the $j$-th
transmitter to the $k$-th receiver is modeled as
\begin{equation}\label{ChanelModel}
 {\utwi{h}}_{k,j}=\left(\frac{200}{d_{k,j}}\right) ^{3.5}l_{k,j}\bar{\utwi{h}}_{k,j}
\end{equation}
where $d_{k,j}$ is the distance  from the $j$-th
transmitter to the $k$-th receiver; $10\log_{10}l_{k,j}\sim {\cal N}(0,8)$ is
a real Gaussian random variable  accounting for the large scale
log-normal shadowing;   $\bar{\utwi{h}}_{k,j}\sim \mathcal{N}_{c}(\utwi{0}_{M},\utwi{I}_{M})$ is a circularly symmetric complex
Gaussian random vector accounting for Rayleigh fast fading.

We define
\begin{equation}
\kappa_{k,j}^{m}\triangleq\frac{{\rm Var}({\hat{h}_{k,j}^{m}})} {{\rm Var}(\delta_{k,j}^{m})} \times 100\%
\end{equation}
where ${\rm Var}({\hat{h}_{k,j}^{m}})$ and ${\rm Var}(\delta_{k,j}^{m})$ are the standard deviations of
the channel $ {h}_{k,j}^{m}$ and
the channel error $\delta_{k,j}^{m}$, respectively. For simplicity,  we assume $\kappa_{k,j}^{m}=\kappa, \forall k,j,m$,
and consider cases of  $\kappa \in\{ 1\%, 5\%, 10\%,  15\% \}$ in the following simulations.
Moreover,  we assume that all receivers have the same SINR level $\alpha$.

%

\begin{figure}[htbp]
 \centering
 \includegraphics[scale=0.8,bb=100 493 480 760]{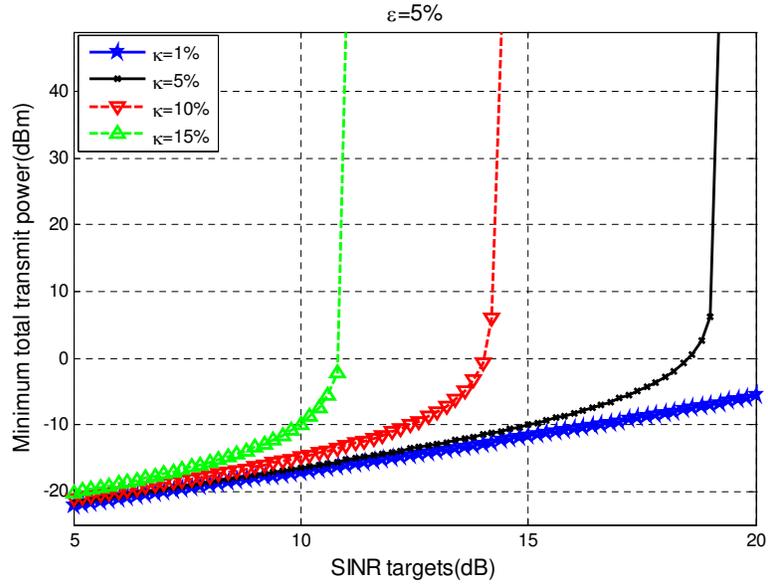}
 \caption{Minimum total transmit power versus SINR level $\alpha$, for different
values of $\kappa$ and for $\varepsilon=5\%$ in a MISO interference channel.}
 \label{PowerMin1}
 \end{figure}

We consider the SINR-constrained power minimization problem in MISO interference channels, given by
(\ref{MinPower}). We solve this problem by using the Bernstein approximation and the LLBCP algorithm, as discussed in Section III.A-B.
First we consider the impact of channel error variance on the power control performance.
Fig. \ref{PowerMin1} shows the minimum total transmit power, $P_{T}^{\min}(\alpha)=\sum_{k} p_{k}(\alpha)$
versus the required SINR level $\alpha$ for the fixed outage probability $\varepsilon=5\%$ and for the different channel uncertainty levels $\kappa$.
It is seen that  when the channel uncertainty   increases,
it takes more power to meet the SINR outage requirement.
For a fixed channel uncertainty level $\kappa$, as the target
SINR value $\alpha$ increases,
  it becomes exceedingly difficult to meet the outage requirement;
  and moreover, the transmit power increases
drastically near some limiting SINR value. This limiting value
is the one which makes the optimization problem infeasible.
Therefore the effect of imperfect CSI is more difficult to cope with when target SINR is high.
As the channel uncertainty   increases, the maximum
feasible SINR value   $\alpha$ also decreases.

\begin{figure}[htbp]
 \centering
 \includegraphics[scale=0.8,bb=100 500 480 760]{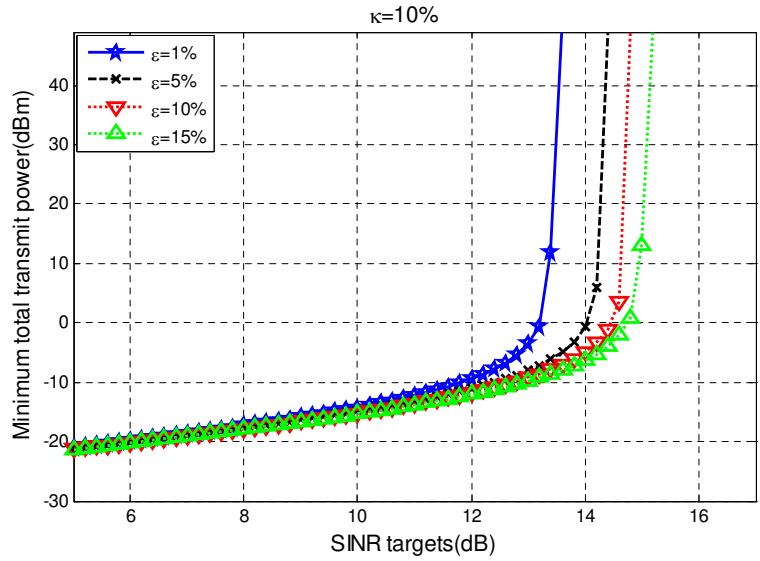}
 \caption{Minimum total transmit power versus SINR level $\alpha$, for different
values of $\varepsilon$ and for $\kappa=10\%$ in a MISO interference channel.}
 \label{PowerMin2}
 \end{figure}

We next consider the impact of the outage probability requirement  on the power control performance.
Fig. \ref{PowerMin2}  illustrates $P_{T}^{\min}(\alpha)$ versus  the target SINR level $\alpha$ for
the fixed channel uncertainty $\kappa=10\%$ and for different outage probability values $\varepsilon$.
It is seen that as the outage requirement becomes more stringent, i.e., when $\varepsilon$ becomes smaller,
it takes more power to meet the SINR outage requirement, and  the maximum
feasible SINR value  $\alpha$ becomes smaller.

\subsection {Max-Min SINR Optimization in MISO Interference Channels}

\begin{figure}[htbp]
 \centering
 \includegraphics[scale=0.8,bb=100 473 480 760]{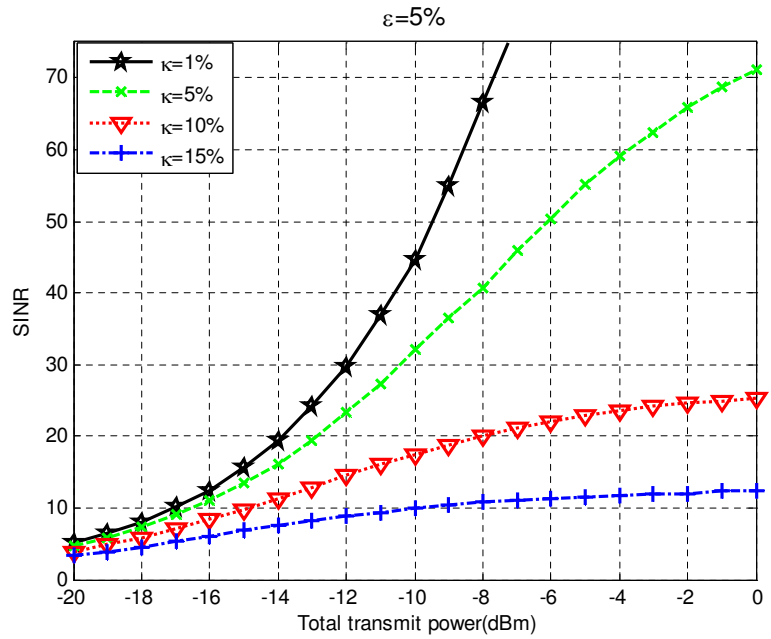}
 \caption{Maximum achievable SINR   versus   maximum allowable total
transmit power, for different
values of $\kappa$ and for $\varepsilon=5\%$, in a MISO interference channel.}
 \label{SINRMax1}
 \end{figure}

\begin{figure}[htbp]
 \centering
 \includegraphics[scale=0.8,bb=100 500 480 760]{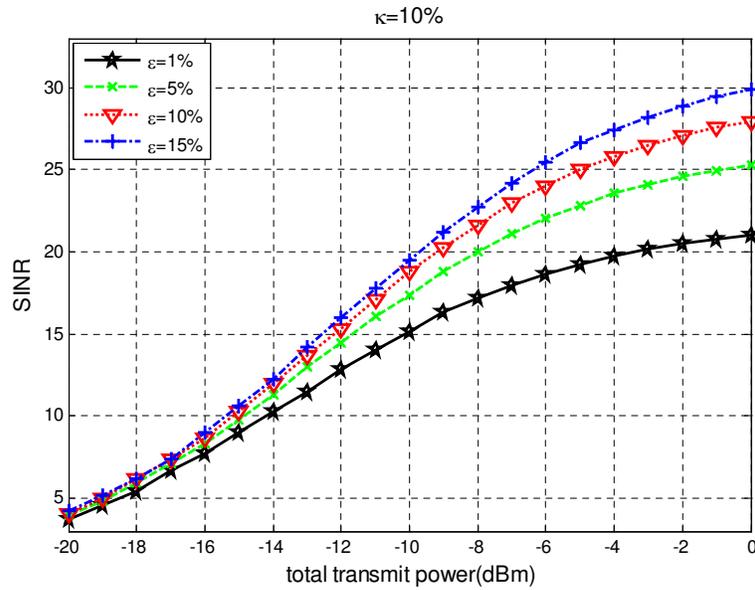}
 \caption{Maximum achievable SINR  versus   maximum allowable total
transmit power, for different
values of $\varepsilon$ and for $\kappa=10\%$, in a MISO interference channel.}
 \label{SINRMax2}
 \end{figure}

We now consider the max-min SINR optimization problems in MISO interference channels under
either individual or total transmit power constraint, given by (\ref{MaxMinRate})-(\ref{MaxMinRatetotal}).
Again the Berinstein approximation and the   LLBCP algorithm are employed to solve the problems, as
outlined in Section III.C.
In Fig. \ref{SINRMax1}, we plot the maximum achievable SINR  versus the maximum allowable total transmit
power for  fixed outage probability $\varepsilon=5\%$ and for   different values of the mismatched error variance $\kappa$. In
Fig. \ref{SINRMax2}, we plot the maximum achievable SINR  versus the maximum allowable total transmit
power
for  fixed $\kappa=10\%$ and for   different   $\varepsilon$.
It is seen that  for a given    maximum allowable total transmit
power, the maximum
achievable SINR  decreases as the channel uncertainty $\kappa$  increases, or as the outage probability $\varepsilon$   decreases.

\begin{figure}[htbp]
 \centering
 \includegraphics[scale=0.8,bb=100 473 480 760]{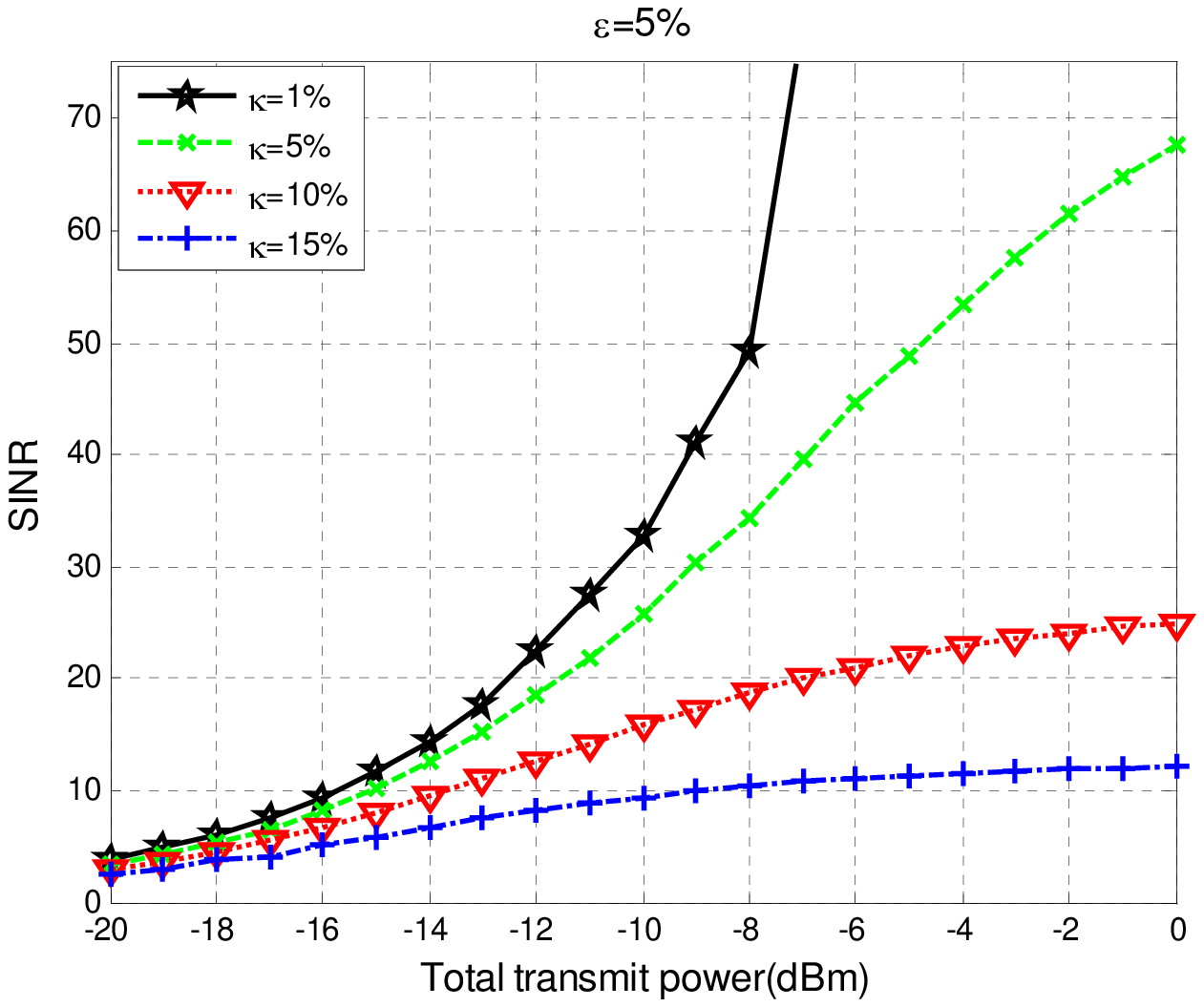}
 \caption{Maximum achievable SINR  versus  maximum allowable total
transmit power, under individual power constraints,  for different
values of $\kappa$ and for $\varepsilon=5\%$, in a MISO interference channel.}
 \label{SINRMax3}
 \end{figure}

\begin{figure}[htbp]
 \centering
 \includegraphics[scale=0.8,bb=100 473 480 760]{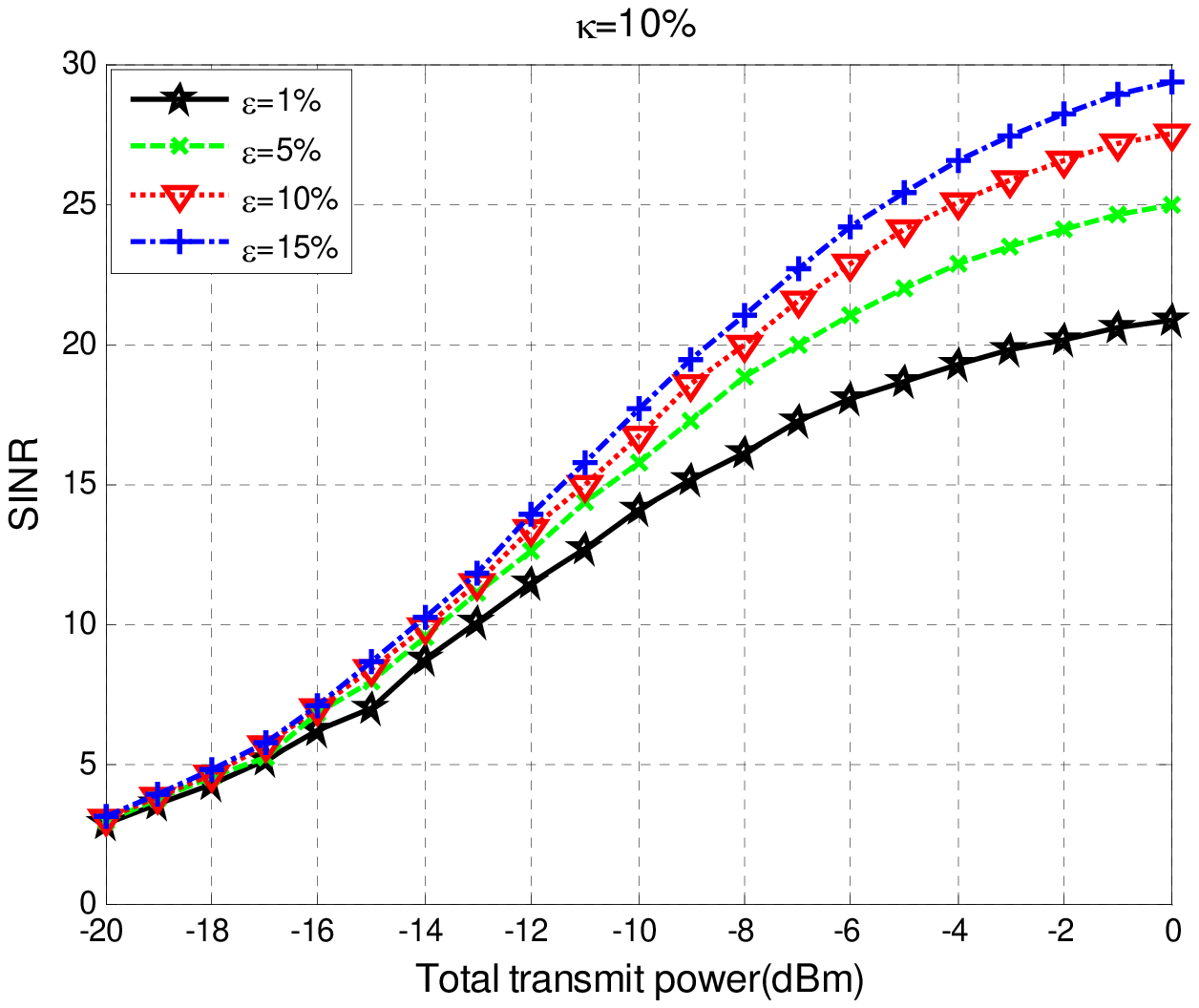}
 \caption{Maximum achievable SINR  versus   maximum allowable total
transmit power, under individual power constraints,  for different
values of $\varepsilon$ and for $\kappa=10\%$, in a MISO interference channel.}
 \label{SINRMax4}
 \end{figure}

Next we consider the case where
the individual   transmit powers are constrained.
We assume that the transmitters have the same maximum allowable
individual transmit power
$\bar{p}_k=\bar{p}_{total}/K$, where $\bar{p}_{total}$ denotes the total transmit power.
Fig. \ref{SINRMax3} shows the maximum achievable
SINR  versus the total transmit power,
for fixed $\varepsilon=5\%$ and for   different values of $\kappa$.
Fig. \ref{SINRMax4} illustrates the maximum achievable SINRs versus the total transmit power,
for   fixed $\kappa=10\%$ and for   different outage probability $\varepsilon$.
It is seen that  the maximum achievable
SINR under the individual transmit power constraint
suffers an SINR loss
as compared to that under the total transmit power constraint.
%
%
%

\subsection {MSE-constrained Power Minimization in MISO Broadcast Channels}

In this section, we illustrate the performance of the proposed Bernstein approximation approach to the MSE-constrained power minimization
in  a downlink MISO system, as discussed in Section IV,   with comparison with the VPI-based approach  proposed in \cite{VucicChance}.

We consider a downlink MISO system with  $K=3$ users and  the basestation is equipped with $M=3$ transmit antennas.
We set a same MSE target for all users and choose $\mu_{1}=\mu_{2}=\mu_{3}=\mu$  from -15dB to -5dB.
The  channel coefficients are generated as i.i.d. complex Gaussian random variables
with zero mean and unit variance. The channel error matrix is set as
$\utwi{\Delta}[k,:]=\sigma_{k}\utwi{\overline{\Delta}}[k,:], k=1, ... ,K$, where $\utwi{\overline{\Delta}}[k,:]
\sim {\cal N}_c( \utwi{0}_M, \utwi{I}_M)$.

\begin{figure}[htbp]
 \centering
 \includegraphics[scale=0.8,bb=100 473 480 760]{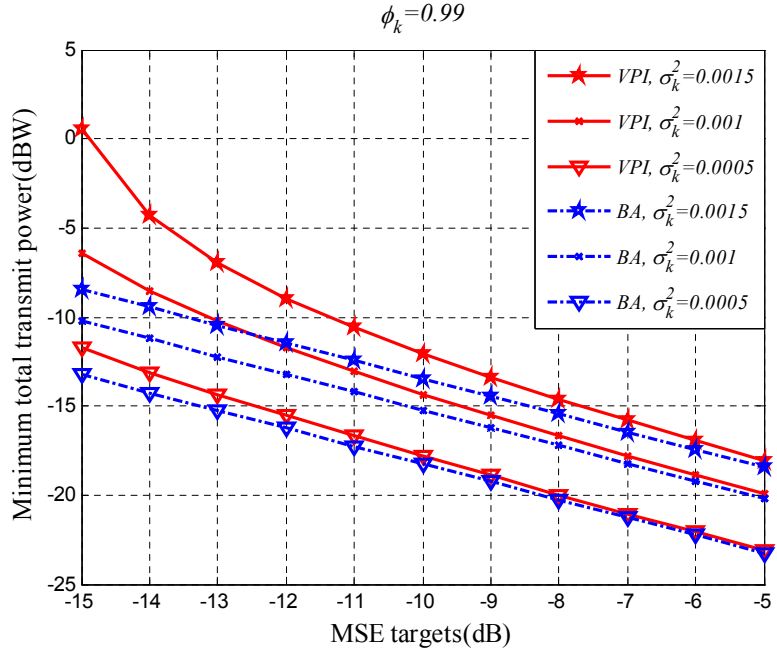}
 \caption{Minimum total  transmit power versus the target MSE, for different
values of $\sigma_{k}^{2}$ and for $\phi=0.99$, in a MISO broadcast channel.}
 \label{MSE_1}
 \end{figure}

First, we compare the total transmit power   of the Bernstein approximation approach and that of the VPI-based method.
In Fig. \ref{MSE_1}, the minimum total transmit power against the target MSE   is plotted,
for three channel error values $\sigma_{k}^{2}$=$1.5\times10^{-3}$, $\sigma_{k}^{2}$=$10^{-3}$ and $\sigma_{k}^{2}$=$0.5\times10^{-3}$,
at the probabilistic guarantee $\phi_{k}=0.99$.
It is seen that the proposed Bernstein approximation approach outperforms the VPI-based method  in that
it achieves lower total transmit power in all cases.
Additional simulations show that the
 Bernstein approximation approach significantly outperforms the VPI-based method when
 the channel uncertainly is high (i.e.,
larger $\sigma_{k}^{2}$), and/or the  targe MSE is low (i.e., small $\mu_k$),
and/or the probabilistic guarantee is stringent (i.e., high $\phi_k$).
For instance, we can see from Fig. \ref{MSE_1} that for $\sigma_{k}^{2}$=$1.5\times10^{-3}$, $\mu=$-15dB, and $\phi_{k}$=0.99, the transmit power difference between the two approaches is about 7dBW.

\begin{figure}[htbp]
\centering
\subfigure[Bernstein approximation approach]
{\label{MSEAA8}
\includegraphics[scale=2.1,bb=100 699 258 764]{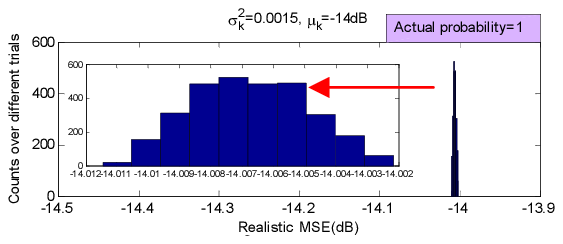}}
\subfigure[VPI based approach]
{\label{MSEAA9}
\includegraphics[scale=0.82,bb=130 590 502 755]{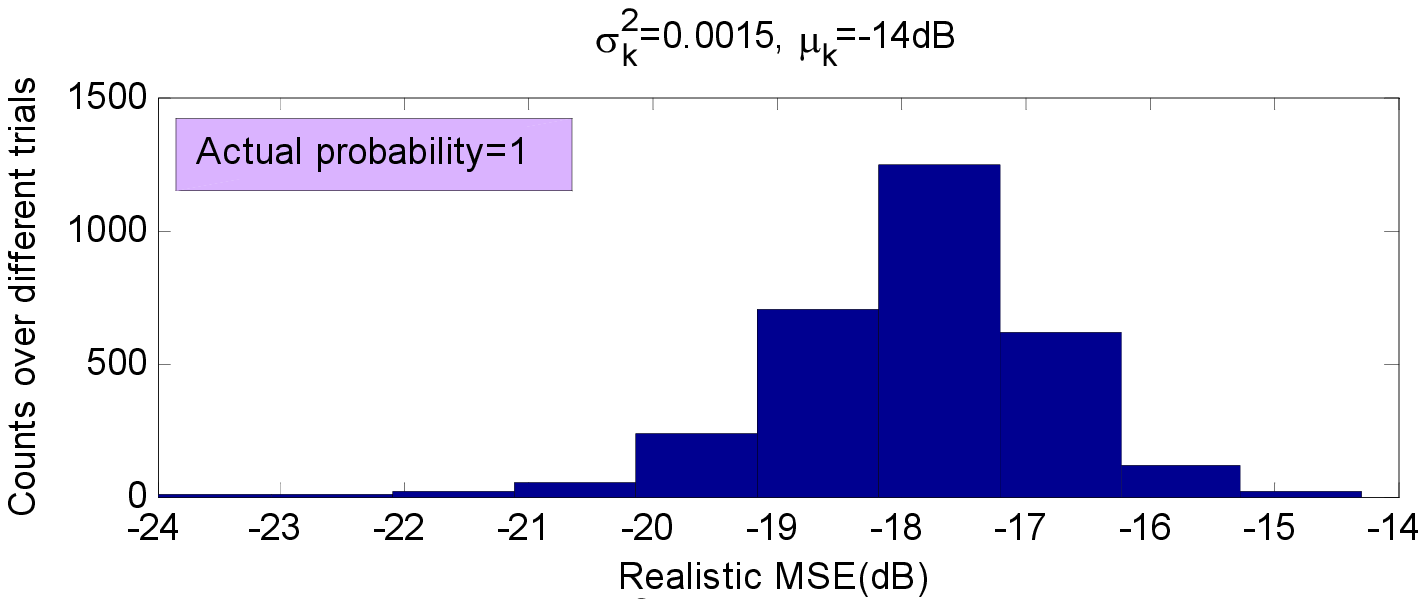}}
\caption{Histograms of the MSE under the Bernstein approximation approach and the VPI-based method.
$\mu_{k}=-14$dB, and
  $\phi_{k}=0.99.$}
\end{figure}

In order to get more insight into the behavior of these two approaches, we verify the actual
probabilistic guarantees by plotting  the histograms of the MSE
in Fig. \ref{MSEAA8}-\ref{MSEAA9}. The system parameters are
  $\sigma_{k}^{2}=1.5\times10^{-3}$,
  $\mu_{k}=-14$dB, and
  $\phi_{k}=0.99.$
The results show that under both approaches, the probabilistic constraints are met. In fact,
the actual probability is larger than the target probability, corroborating  that both approaches
provide conservative approximations to the original probabilistic constraints.
However, it is seen that the histogram of the VPI-based method is much more spread than
that of the Bernstein approximation approach. In particular, the MSE realizations under the
Bernstein approximation approach concentrate sharply around the target MSE value $\mu$,
whereas the MSE realizations under the VPI-based approach spread around an MSE value that is
much lower than the target value $\mu$. Hence the
  VPI-based method provides  a more conservative approximation to the original probabilistic constraint,
  and thus  needs more transmit power to meet the resulting more stringent MSE constraint than the actual
  target constraint.


\section{Conclusions}

We have treated the problems of robust power allocation in multiuser MISO
systems with imperfect transmitter-side CSI. The
multi-antenna transmitters are assumed to employ some fixed beamformers to
transmit data and the transmit powers need to be optimized to satisfy
certain QoS constraints, taking into account the uncertainty in the available
CSI. Specifically, for MISO interference channels, we have considered the
transmit power minimization problem and the max-min SINR problem, subject
to the constraints on the SINR outage probabilities. For MISO broadcast
channels, we have considered the transmit power minimization problem
subject to the constraints on the MSE outage probabilities. Our key
contribution is to employ the Bernstein approximation to conservatively
transform the probabilistic constraints into deterministic ones, and
consequently convert the original stochastic optimization problems into
convex optimization problems. We have provided extensive simulation
results to demonstrate the effectiveness of the proposed robust power
optimization techniques.

\section*{Appendix}
\subsection{Proof of Proposition 1}
\noindent \textbf{Proof:}
We have
\begin{equation}\label{appendix1}
{\Gamma _k} = \frac{{{p_k}{{\left| {{\utwi{h}_{kk}^H}{\utwi{g}_k}} \right|}^2}}}
{{\eta _k^2 + \sum\limits_{j \ne k} {{p_j}{{\left| {{\utwi{h}_{kj}^H}{\utwi{g}_j}} \right|}^2}} }} \leqslant {\alpha _k}
\end{equation}
\begin{equation}\label{appendix2}
\Longleftrightarrow {\alpha _k}\eta _k^2 + {\alpha _k}\sum\limits_{j \ne k} {{p_j}{{\left| {{\utwi{h}_{kj}^H}{\utwi{g}_j}} \right|}^2}}  - {p_k}{\left| {{\utwi{h}_{kk}^H}{\utwi{g}_k}} \right|^2} \geqslant 0
\end{equation}
Plugging (\ref{Equ2}) into (\ref{appendix2}), we obtain (\ref{appendix3}),
\begin{eqnarray}\label{appendix3}
{F_k}\left( {{\utwi{p}},{{\utwi{\xi }}_k}} \right) &\triangleq& {{\alpha _k}\eta _k^2 + {\alpha _k}\sum\limits_{j \ne k} {{{p_j}\underbrace {{{\left| {\utwi{\hat {h}}_{kj}^H\utwi{g}_j + \utwi{\delta}_{kj}^H\utwi{g}_j} \right|}^2}}_{{\xi}_{kj}}} }}
 -  {{p_k}\underbrace {{{\left| {\utwi{\hat {h}}_{kk}^H\textbf{g}_k + \utwi{\delta} _{kk}^H\utwi{g}_k} \right|}^2}}_{{\xi} _{kk}}} \nonumber \\
 &=&  {{\alpha _k}\eta _k^2 + {\alpha _k}\sum\limits_{j \ne k} {{{p_j} {{\xi}_{kj}}} }}
 -  {{p_k} {{\xi} _{kk}}}  \geqslant 0
\end{eqnarray}
where ${\boldsymbol\xi_k} \triangleq \left[ {{\xi _{k1}},{\xi _{k2}},...,{\xi _{kK}}} \right]^T$.
Note that ${F_k}\left( {{\utwi{p}},{{\utwi{\xi }}_k}} \right)$ in (\ref{appendix3}) is in the form of (\ref{BerTheorem1}).

Since $\utwi{\delta} _{kj}, \forall j$ are independent, $\xi _{kj}, \forall j$ are independent  random variables.
Furthermore, from (\ref{delta}),  it follows that
$\utwi{\delta} _{kj}^H\utwi{g}_j + \utwi{\hat {h}}_{kj}^H\utwi{g}_j \stackrel{\mbox{\small {i.i.d.}}}{\sim} \mathcal{N}_c\left( {\utwi{\hat {h}}_{kj}^H\utwi{g}_j, \sigma^2_{kj}{{\left| {\boldsymbol{1}^T\utwi{g}_j} \right|}^2}} \right)$.  Then by normalizing $\xi_{kj}$ by the variance $\frac {\sigma^2_{kj}} 2 \left| {\boldsymbol{1}^T\utwi{g}_j} \right|^2$ of the real or imaginary component, we obtain the following noncentral
$\chi^2$ random variable with two degrees of freedom
\begin{eqnarray}
\frac {\xi_{kj}} {\frac {\sigma^2_{kj}} 2 \left| {\boldsymbol{1}^T\utwi{g}_j} \right|^2 }
 &\sim&  \chi_2^2 \Big(
 \frac { | \utwi{\hat {h}}_{kj}^H\utwi{g}_j   |^2 }
 {\frac {\sigma^2_{kj}} 2  | {\boldsymbol{1}^T\utwi{g}_j}  |^2 }
 \Big).
\end{eqnarray}
Thus can write the logarithm of the moment generating function of ${F_k}\left( {{\utwi{p}},{\boldsymbol\xi_k}} \right)$ as
(\ref{MomentGenerating})
  \begin{eqnarray}\label{MomentGenerating}
 &&\log\mathbb{E} \left[ {\exp \left( {{t_{k}^{ - 1}}{F_k}\left( {{\utwi{p}},{\boldsymbol\xi_k}} \right)} \right)} \right]
   = \log{\mathbb E}\left[ {\exp \left(  {t_{k}^{ - 1}}{\alpha _k}\eta _k^2 + {t_{k}^{ - 1}}{\alpha _k}\sum\limits_{j \ne k} { {{p_j} \xi_{kj}} }  - {{t_{k}^{ - 1}}{p_k} \xi_{kk} } \right)} \right] \notag \\
   &=&   {{t_{k}^{ - 1}}{\alpha _k}\eta _k^2}  + \sum\limits_{j \ne k}  \log{\mathbb{E}}\Big\{\exp \Big[ \underbrace {\Big(\frac {\sigma^2_{kj}} 2
     | {\boldsymbol{1}^T \utwi{g}_j}  |^2   t_{k}^{ - 1}  \alpha _k  p_j\Big)}_t\underbrace {\Big(\frac {\xi_{kj} }{\frac {\sigma_{kj}^2} 2
   { | {\boldsymbol{1}^T \utwi{g}_j}  |}^2} \Big)}_X \Big] \Big\}\notag \\
  && +  \log{\mathbb{E}}\Big\{\exp \Big[   {-\Big(\frac {\sigma^2_{kk}} 2
     | {\boldsymbol{1}^T \utwi{g}_k}  |^2   t_{k}^{ - 1}     p_k\Big)}  {\Big(\frac {\xi_{kk} }{\frac {\sigma_{kk}^2} 2
   { | {\boldsymbol{1}^T \utwi{g}_k}  |}^2} \Big)}  \Big] \Big\} \notag\\
   &=& {{t_{k}^{ - 1}}{\alpha _k}\eta _k^2} + \sum\limits_{j \ne k} { {\left[ {\frac{{t_k^{ - 1}{\alpha _k}{p_j}{{\left| {\utwi{\hat {h}}_{kj}^H \utwi{g}_j } \right|}^2}}}
{{1 - t_k^{ - 1}{\alpha _k}{p_j} {\sigma_{kj}^2} {{\left| {\boldsymbol{1}^T \utwi{g}_j } \right|}^2}}} - \log \left( {1 - t_k^{ - 1}{\alpha _k}{p_j}\sigma _{kj}^2{{\left| {\boldsymbol{1}^T \utwi{g}_j } \right|}^2}} \right)} \right]} } \notag \\
&&- { {\frac{{{t_k^{-1}p_k}{{\left| {\utwi{\hat {h}}_{kk}^H \utwi{g}_k } \right|}^2}}}
{{1 +  t_k^{ - 1}{p_k}\sigma_{kk}^2{{\left| {\boldsymbol{1}^T \utwi{g}_k } \right|}^2}}} -   \log \left( {1 +  t_k^{ - 1}{p_k}
\sigma_{kk}^2{{\left| {\boldsymbol{1}^T \utwi{g}_k} \right|}^2}} \right)} }
\end{eqnarray}
where (\ref{MomentGenerating}) follows from the fact that for
 $X \sim \chi _2^2\left( \lambda  \right)$, we have
\begin{equation}
{\mathbb{E}}\left\{ {\exp \left( {tX} \right)} \right\} = \frac{{\exp \left( {\frac{{\lambda t}}
{{1 - 2t}}} \right)}}
 {\left( {1 - 2t} \right)}, \ \ \ {\rm with}\ \  t < \frac 1 2. \label{ttt.eq}
 \end{equation}

 Now using (9), we obtain the Bernstein approximation (\ref{appendix4}) to the probabilistic constraint
 $\Pr \left( {{\Gamma _k} \leqslant \alpha_k} \right) \leqslant {\varepsilon _k}$ in (\ref{MaxMinRate}).
\begin{eqnarray}\label{appendix4}
 {\inf }_{{t_k} >\rho_k }  {G_k}\left( {{\utwi{p}},{t_k}} \right)
  \triangleq t_k \log\mathbb{E} \left[ {\exp \left( {{t_{k}^{ - 1}}{F_k}\left( {{\utwi{p}},{\boldsymbol\xi_k}} \right)} \right)} \right]
  - t_k \log\varepsilon _k \
 \leqslant 0. \label{tt.eq}
 \end{eqnarray}
 Substituting (\ref{MomentGenerating}) into (\ref{tt.eq}), we obtain (\ref{Gk1}).
Note that in order to meet the condition $t<\frac 1 2$ in (\ref{ttt.eq}), we should have
 \begin{eqnarray}
\;\; \frac {\sigma^2_{kj}} 2
     | {\boldsymbol{1}^T \utwi{g}_j}  |^2   t_{k}^{ - 1}  \alpha _k  p_j  &<& \frac 1 2, \ \ \ j\neq k
     \end{eqnarray}
 Hence we have
 \begin{eqnarray}
 t_k > \alpha_k \max_{ j \neq k}  \{   \sigma^2_{kj}p_j | {\boldsymbol{1}^T \utwi{g}_j}  |^2   \} \triangleq \rho_k.
 \end{eqnarray}
 \ \hfill $\Box$

\subsection{Proof of Proposition 2}
The proof follows the similar line as that in \cite{Tajer}.
Let us denote the set of powers obtained from solving $\mathcal{S}({\utwi{\bar p}})$ by $\left\{ {p_k^*} \right\}$ and their corresponding minimal SINR by $\alpha_k^*$.  From the definition of $S({\utwi{\bar p}})$ we have ${p_k} \leqslant {\bar p_k},\forall k$ and $\mathop {\min }\limits_k \left\{ {\alpha_k^*} \right\} = \mathcal{S}({\utwi{\bar p}}) \Rightarrow \alpha_k^* \geqslant \mathcal{S}({\utwi{\bar p}}),\forall k$. As a result from the definition of $\mathcal{P}({\utwi{\bar p}},a)$ we find that for the choice of $\left\{ {p_k^*} \right\}$, the choice of $b=1$ is achievable for $\mathcal{P}({\utwi{\bar p}},\mathcal{S}({\utwi{\bar p}}))$ and therefore $\mathcal{P}({\utwi{\bar p}},\mathcal{S}({\utwi{\bar p}})) \leqslant 1$.

Next we show that $\mathcal{P}({\utwi{\bar p}},\mathcal{S}({\utwi{\bar p}}))$ cannot be less than one. Let us denote the set of powers obtained by solving $\mathcal{P}({\utwi{\bar p}},\mathcal{S}({\utwi{\bar p}}))$ by $\left\{ {p_k^{**}} \right\}$. From the definition of $\mathcal{P}({\utwi{\bar p}},\mathcal{S}({\utwi{\bar p}}))$ we clearly have $\alpha_k^{**} \geqslant \mathcal{S}({\utwi{\bar p}}),\forall k$. If $\mathcal{P}({\utwi{\bar p}},\mathcal{S}({\utwi{\bar p}})) < 1$, i.e., if
$\mathop {\max }\limits_k \frac{{p_k^{**}}}{{{{\bar p}_k}}} = c < 1$, then we define the set of powers $\left\{ {{{\hat p}_k}} \right\} = \left\{ {{{p_k^{**}} \mathord{\left/{\vphantom {{p_k^{**}} c}} \right.\kern-\nulldelimiterspace} c}} \right\}$.
$\left\{ {{{\hat p}_k}} \right\}$ clearly satisfy the power constraints and we have their corresponding SINRs satisfying
\[
\begin{array}{l}
 \hat \Gamma _k  = \frac{{\hat p_k \left| {\utwi{h}_{kk} \utwi{g}_k } \right|^2 }}{{\sigma _k^2  + \sum\limits_{j \ne k} {\hat p_j \left| {\utwi{h}_{kj} \utwi{g}_j } \right|^2 } }} = \frac{{\frac{{p_k^{**} }}{c}\left| {\utwi{h}_{kk} \utwi{g}_k } \right|^2 }}{{\sigma _k^2  + \sum\limits_{j \ne k} {\frac{{p_j^{**} }}{c}\left| {\utwi{h}_{kj} \utwi{g}_j } \right|^2 } }} \\
\;\;\;\;\; = \frac{{p_k^{**} \left| {\utwi{h}_{kk} \utwi{g}_k } \right|^2 }}{{c\sigma _k^2  + \sum\limits_{j \ne k} {p_j^{**} \left| {\utwi{h}_{kj} \utwi{g}_j } \right|^2 } }} > \frac{{p_k^{**} \left| {\utwi{h}_{kk} \utwi{g}_k } \right|^2 }}{{\sigma _k^2  + \sum\limits_{j \ne k} {p_j^{**} \left| {\utwi{h}_{kj} \utwi{g}_j } \right|^2 } }} \\
 \end{array}
\]
Since $c<1$, ${\hat \Gamma _k} > \Gamma _k^{**}$. Therefore, we have found a set of powers which satisfy the power constraints and yet yield a strictly larger minimal SINR compared to what the powers $\left\{ {p_k^{**}} \right\}$ obtain. This contradicts the optimality of $\left\{ {p_k^{**}} \right\}$ and therefore $\mathcal{P}({\utwi{\bar p}},\mathcal{S}({\utwi{\bar p}})) = 1$.
The strict monotonicity and continuity of $\mathcal{P}({\utwi{\bar p}},a)$ in $a$, at any strictly feasible region, follows from a similar line of argument.
\hfill$\Box$

\subsection{Proof of Proposition 3}
\noindent \textbf{Proof:}
To be consistent with (\ref{Ber3B}), we rewrite the constraint $\mathrm{Pr} \left(\mathrm{MSE}_{k}\leq \mu_k \right) \geq \phi_{k}$
as $\mathrm{Pr} \left(\mathrm{MSE}_{k}\geq \mu_k \right) \leq 1-\phi_{k}.$
Using (\ref{ttt1.eq}) and (\ref{ttt2.eq}) we have
\begin{equation}
\mathrm{MSE}_k  = q_k^{ - 1} \utwi{\Delta}[k,:] \utwi{GQG}^H \utwi{\Delta}[k,:]^H  + q_k^{ - 1} \eta_k^2.
\end{equation}
Thus  the condition $\mathrm{MSE}_{k}\geq \mu_k$ becomes
\begin{eqnarray}\label{MSE1}
F_k( \utwi{Q}, \utwi{\Delta}[k,:] ) \triangleq    \utwi{\Delta}[k,:] \utwi{GQG}^H \utwi{\Delta}[k,:]^H  + \eta _k^2  - q_k \mu _k
   \ge 0.
\end{eqnarray}
 Note that $\utwi{\Delta}[k,:]$ is a zero-mean complex Gaussian random vector with a diagonal covariance matrix  $\utwi{\Lambda}_k
 \triangleq {\rm diag}( \sigma_{k,1}^2, ..., \sigma_{k,M}^2)$.
Thus we have (\ref{appendix5}) \cite{Mtx_book}.
\begin{eqnarray}\label{appendix5}
\mathbb{E}\Big\{\exp\Big(t_k^{-1}\utwi{\Delta}[k,:] \utwi{GQG}^{H} \utwi{\Delta}[k,:]^{H}\Big)\Big\}=\frac{1}{\sqrt{\det(\utwi{I}-\frac{2}{ {t}_k}\utwi{\Lambda}_k \utwi{GQG}^{H} )}}.
\end{eqnarray}
Then the Bernstein approximation to the probabilistic constraint $\mathrm{Pr} \left(\mathrm{MSE}_{k}\geq \mu_k \right) \leq 1-\phi_{k}$
becomes $\inf_{t_k >0} G_k^{\mathrm{MSE}} \left( {\utwi{Q}, {t}_k } \right) \leq 0$, where $G_k^{\mathrm{MSE}} \left( {\utwi{Q}, {t}_k } \right)$ is defined in (\ref{GkMSE2})
\begin{eqnarray}\label{GkMSE2}
 G_k^{\mathrm{MSE}} \left( {\utwi{Q}, {t}_k } \right) &\triangleq &
  t_k \log\mathbb{E} \left[ {\exp \left( {{t_{k}^{ - 1}}{F_k}\left(  {\utwi{Q}},{\boldsymbol\Delta}[k,:]  \right)} \right)} \right]
  - t_k \log(1-\phi_k)  \nonumber \\
 &=&  (\eta_k^2  - q_k \mu _k)
 - \frac{ {t}_k}{2} \log \det\Big(\utwi{I}-\frac{2}{ {t}_k}\mathbf{\Lambda}_k \utwi{GQG}^{H} \Big)-  t_k \log(1-\phi_k).
\end{eqnarray}
\ \hfill  $\Box$


\end{document}